\documentstyle[12pt,titlepage]{article}

\newcommand{\beq}{\begin{equation}}
\newcommand{\eeq}{\end{equation}}
\newcommand{\beqa}{\begin{eqnarray}}
\newcommand{\eeqa}{\end{eqnarray}}

\begin{document}

\sloppy

\title{The Supercooling of a Nematic Liquid Crystal}

\author{
P. De*\\
Department of Physics\\
University of Rhode Island\\
Kingston, RI  02881\\ \\ \\
Robert A. Pelcovits, \\
 Department of Physics,\\
Brown University\\
Providence, RI   02912\\ \\ \\
E. Vogel and  J. Vogel\\
Institut f\"ur Theoretische Physik, \\
R.W.T.H. Aachen \\
Templergraben 55 \\
W-5100 Aachen, Germany \\ \\ \\ }

\maketitle
%\setcounter{page}{1}
%\input{pial.6}
%\input{pial.7}
%\input{pial.8}
%\input{pial.REF}
%\input{pial.Cap}
%\input{pial.Tab}
%\end{document}
\section*{Abstract}
We investigate the supercooling of a nematic liquid crystal using
fluctuating non-linear hydrodynamic equations. The Martin-Siggia-Rose
formalism is used to calculate renormalized transport coefficients to
one-loop order. Similar theories for isotropic liquids have shown
substantial increases of the viscosities as the liquid is supercooled
or compressed due to feedback from the density fluctuations which are
freezing. We find similar results here for the longitudinal and
various shear viscosities of the nematic. However, the two viscosities
associated with the nematic director motion do not grow in any
dramatic way; i.e.\ there is no apparent freezing of the director modes
within this hydrodynamic formalism. Instead a glassy state of the
nematic may arise from a  ``random anisotropy" coupling of the
director to the frozen density.

\setcounter{equation}{0}
\section{Introduction}
The study of a supercooled nematic liquid crystal and the possible
formation of a nematic glass$^1$ is potentially richer than corresponding
studies of supercooled simple fluids. The presence of anisotropy in
the nematic liquid due to the overall alignment of the molecular long
axes introduces orientational degrees of freedom into the description
of the system and yields a model similar to a spin glass with
translational degrees of freedom coupled to the spin fluctuations. We
use the term ``nematic glass" to describe a liquid crystal where both
the translational (density) and orientational (director) fluctuations
are frozen. Assuming that we supercool the liquid starting from its
nematic phase (rather than the isotropic phase) we expect a nematic
glass to have long-range orientational order and thus be similar to a
mixed  magnetic phase where both ferromagnetic and spin glass
order coexist.$^2$
The
presence of both translational (density) and orientational (director)
fluctuations and their coupling leads to a novel glass-forming
system.$^3$
Glass formation in this system could potentially occur in a two-stage
process where the density or director modes freeze first, followed by
the other, or in a process where both freeze simultaneously.
\par
In recent years a theoretical approach to the study of glass formation
has been developed using either mode-coupling calculations$^4$ or
fluctuating nonlinear  hydrodynamics.$^5$ These approaches were initiated
by Leutheusser$^4$ who showed that a model of a dense fluid obtained from
kinetic theory exhibits a sharp glass transition where the system
becomes nonergodic. Das, Mazenko, Ramaswamy and Toner$^5$ developed an
equivalent model on the basis of fluctuating nonlinear dynamics.
Subsequently, Das and Mazenko$^6$ discovered a nonhydrodynamic mechanism
which cuts off the sharp transition, leading to a rounded transition.
These authors claim that the cutoff is due to the proper mathematical
treatment of the relationship $\vec{P}=\rho\vec{V}$ where $\vec{P}$ is
the momentum density, $\rho$ is the mass density, and $\vec{V}$ is the
velocity field. The physical origin of this cutoff is unknown. More
recently, Schmitz et al.$^7$ have argued that the calculation of Das and
Mazenko is not correct. Furthermore, they claim that the perturbative
calculations of refs. 5 and 6 do not properly account for detailed balance.
Restoring detailed balance to the perturbation theory apparently
restores ergodicity and leads to a rounded transition. However, Das
and Mazenko and Schmitz et al. all agree that the original Leutheusser
theory as well as its hydrodynamic version with no cutoff mechanism
are a good approximation to the more complete theories of refs. 6 and
7 in describing the growth of the viscosities in the pre-glass transition
regime. Eventually the growth of the viscosities is limited by one of
these cutoff mechanisms.
Comparison of these theories with
experiments yields some encouraging agreement,$^8$ though this agreement
is by no means complete,  especially at very low frequencies.
\par
In this paper we study the formation of a nematic glass using
fluctuating nonlinear hydrodynamics.  We will limit our attention
primarily to the pre-glass transition regime in light of the
discussion above, and we will not concern ourselves with the question
of detailed balance or the Das Mazenko cutoff mechanism.
The primary advantage of the fluctuating hydrodynamic
theory is that new slow variables (such as the director modes or
variables associated with broken translational symmetries$^9$) are readily
incorporated. As described in detail in the next section, we
supplement the nonlinear hydrodynamics equations used by Das et al. to
study simple fluids, with two equations describing the dynamics of the
director in a compressible nematic. The full set of equations we
employ also includes a nonlinear coupling
$(\hat{n}\cdot\vec{\bigtriangledown}\rho)^{2}$
between the density and director $\hat{n}$.
Density fluctuations couple to some, but not all of the nematic
viscosities through this coupling. The Leutheusser feedback mechanism
then leads to the enhancement of these viscosities with a universal
prediction for their power law behavior. However, there is apparently
no director feedback mechanism within this  formalism, and we find no
evidence for  the freezing of the director modes. Nevertheless, it is
tempting to speculate that the term
($\hat{n}\cdot\vec{\bigtriangledown}\delta\rho)^2$ in the free energy
leads to freezing of the director. Once the density has frozen, or
almost frozen  this
term will mimic a random anisotropy field in an amorphous
magnet.$^{10}$ The
frozen density gradients $\vec{\bigtriangledown}\rho$ play the role
of the quenched random axis. This model is believed to exhibit a spin-
glass phase, which would correspond in our case to frozen director
modes, with \underbar{no} nematic long range order in an infinite
system. However, for a finite sized system, there would be apparent
long-range order, especially if the coefficient of the biquadratic
term is small.  Whether this proposed freezing of the director takes
place immediately upon freezing of the density or requires further
supercooling is unclear and beyond the scope of our present
theoretical treatment.
\par
This paper is organized as follows: in the next section we formulate
the nonlinear hydrodynamic equations for a compressible nematic
including the above-mentioned nonlinear coupling of density and
director modes. In section III we use the Martin-Siggia-Rose formalism
to study the effects of the nonlinearities on the bare transport
coefficients. Finally in section IV we discuss the implications of our
calculations for the growth of the viscosities and the mode structure
as the nematic is supercooled.
Various technical  details appear in the appendix.

\setcounter{equation}{0}
\section{Non-Linear Hydrodynamic Equations}
In contrast to the spherical molecules of simple liquids, the molecules of
 liquid
crystals are elongated in shape.$^{11}$ Intermolecular interactions cause the
 anisotropic
molecules to align along a preferred direction  denoted by the
director  vector ${\bf n}({\bf x},t)$.
Fluctuations in the director can extend over
macroscopic distances and decay over finite times, leading to new hydrodynamic
modes in addition to the shear and sound wave modes of a simple liquid. The
 broken
rotational symmetry also permits more viscosities: in a compressible
nematic liquid crystal,
there are six independent viscosities.$^{12}$
The linear equations for the hydrodynamic modes of a liquid crystal have been
known for almost twenty years.$^{13}$ A systematic method for deriving
 the nonlinear
contributions is to write them in the form of generalized Langevin
equations:$^{14}$
\beq
%\label{3.1}
\partial \psi_{i}/ \partial t = \bar{V}_{i}[\psi]
-\sum_j \int d^{3}x'\; \Gamma_{ij}({\bf x}') \cdot
\frac{\delta H}{\delta \psi_j({\bf x'})} +
\Theta_{i},
\eeq
where $\psi_{i}({\bf x},t)$
represents one of the seven possible hydrodynamic fields: the mass density
$\rho({\bf x},t)$,
three components of the momentum density ${\bf P}({\bf x},t)$,
the energy density $e({\bf x},t)$
and  two components of the fixed-length director $ {\bf n (\underline
{x}},t)$.  The label $i$ in this equation denotes the type of field,
as well as the vector index on $\vec{P}$ and $\vec{n}$.
$\bar{V}_{i}[\psi]$
represents the reversible part of the dynamic
equations and is given by:
\beq
%\label{3.2}
\bar{V}_{i} [\psi( {\bf x} )] = \sum_j \int d^{3} x' \{\psi_i({\bf x}) ,
 \psi_j({\bf x}')\}
\frac {\delta H}{\delta \psi_j({\bf x}')}.
\eeq
and $H$ is the energy obtained by integrating the free energy density
$F[\psi]$:
\beq
%\label{3.3}
 H = \int d^{3} x F [ \psi({\bf x})].
\eeq
The Poisson bracket in eqn. (2.2) is defined in its usual manner as:
\beq
%\label{3.4}
\{\psi_{i}, \psi_{j}\}  = \sum_{\alpha,\beta,k} [ \frac {\delta \psi_{i} }
                                                 { \delta r_{k}^{\alpha \beta}
}
  \frac {\delta \psi_{j} } {\delta P_{k}^{\alpha \beta} } -
  \frac {\delta \psi_{i} } {\delta P_{k}^{\alpha \beta} }
  \frac {\delta \psi_{j} } {\delta r_{k}^{\alpha \beta} } ]
\eeq
where $r_{k}^{\alpha\beta}$
is the $k^{th}$
component of the vector ${\bf r}^{\alpha\beta}$
which points to the $\beta^{th}$
 atom
of the $\alpha^{th}$
molecule of the liquid crystal.
The second term on the right hand side of the  equation of motion
(2.1)
 represents
the dissipative contributions in form of the dissipative matrix
$ \Gamma_{ij} ({\bf x},{\bf x}')$. Finally, $\Theta_{i} ( {\bf x},t ) $
denotes a Gaussian noise source
which satisfies:
\beq
%\label{3.5}
\langle \Theta_{i} ({\bf x},t) \Theta_{j} ({\bf x}',t')\rangle
=  2 k_B T\: \Gamma_{ij}({\bf x}-{\bf x}') \delta (t-t') .
\eeq
It has been argued that the dominant transport anomalies at the glass
transition
in a simple liquid
are due to the slow decay of density fluctuations, and the  effects of
energy fluctuations can be ignored.$^{4,5}$
We shall denote the global
preferred direction of orientation by ${\bf n}_{0}={\bf e_{z}}$.
Local fluctuations in orientation
are specified by $\delta{\bf n}={\bf n}-{\bf n}_0$
where we restrict $\delta{\bf n}=(\delta n_x,\delta n_y,0)$
to linear order in $\delta n_i.^{15}$ Then ${\bf n}$ satisfies
${\bf n}\cdot {\bf n} = 1$ up to order $(\delta {\bf n})^2 $.
Furthermore one can assume that the wave vector ${\bf k}$ of a
disturbance lies in the $x-z$
plane. This allows us to treat $\delta n_{x}$ and $\delta n_{y}$
as longitudinal and transverse fluctuations with respect to ${\bf k}$
 respectively.$^{16}$ Our set of dynamical variables $\psi_i$ then includes
the density, three components of the momentum, and the director
fluctuations, $n_x$ and $n_y$, which are respectively equal to $\delta
n_x$, and $\delta n_y$ in the present approximation.

In order to evaluate the Poisson bracket between the six dynamical variables
$ \rho, P_x, P_y, P_z, n_x $ and $ n_y $
we need a microscopic description for each variable. We define:
\beq
\rho({\bf x},t) = \sum_{\alpha, \beta} m^\beta \delta
( {\bf x} - {\bf r}^{\alpha \beta} (t) ),
\eeq
\beq P_{i} ({\bf x},t) = \sum_{\alpha, \beta}
P_{i}^{\alpha \beta}(t) \delta ({\bf x}-{\bf r}^{\alpha \beta}(t)) ,
\eeq
 \beq
n_{i}({\bf x},t) =
\frac{1}{\sqrt{N}}
\sum_{\alpha} n_{i}^{\alpha}(t)\delta _{{\bf x},{\bf R}^{\alpha}}.
\eeq
We model an elongated molecule with two atoms only: ${\bf R}^{\alpha}$
is the center of mass of the $\alpha^{th}$
nematic molecule and $n_{i}^{\alpha}~=~ r_{i}^{\alpha}~/~\mid
 {\bf r}^{\alpha} \mid$,
and
${\bf r}^{\alpha}$ is the relative vector between the atoms, which, we
 shall assume, model an
elongated molecule of mass $m^\alpha$.
The variable $\beta$ in ${\bf r}^{\alpha \beta}$
takes on values 1 or 2,
and   ${\bf r}^{\alpha} =  {\bf r}^{\alpha 1} -
 {\bf r}^{\alpha 2} $
while ${\bf R}^{\alpha} = ({\bf r}^{\alpha 1} +
 {\bf r}^{\alpha 2})/2 $.
\par
With the definitions (2.5) -- (2.8), the Poisson brackets required in
(2.2) and defined in (2.4) can be computed in a straightforward manner
with the following results:
\beq
\{ \varrho({\bf x}),P_j({\bf x}') \}  =     \nabla_j (\varrho(x)
    \delta({\bf x}-{\bf x'}));
\eeq
\beq
\{ P_i({\bf x}),P_j({\bf x'})     \}
 =  -\nabla_j
                ( \delta({\bf x}-{\bf x'}  ) P_i({\bf x})) +
             \nabla'_i (\delta({\bf x}- {\bf x'})
              P_j({\bf x'}));
\eeq
\beqa
%\label{3.14}
\{ P_i({\bf x}),n_j({\bf x'})    \} & = &
                                         [
                            (\lambda+1)\delta_{ij} n_k({\bf x'})/2
  +(\lambda-1)\delta_{kj} n_i({\bf x'})/2-\lambda (n_i n_j
 n_k)({\bf x'})
                                         ] \cdot \nonumber \\
& & \nabla_k [\delta({\bf x}-{\bf x}')]
+\delta({\bf x}-{\bf x'}) (\nabla'_i n_j({\bf x'})).
\eeqa
where $\nabla_{i} \equiv \partial /\partial x_{i} $
and   $\nabla_{i}'\equiv \partial /\partial x_{i}'$
etc.   In writing eqn. (2.10) we have introduced the ``form  factor"
$\lambda$ which is related to the shape of the molecules and equals
unity only in the limit of infinitesimally thin molecules.$^{17}$
All other Poisson brackets are zero.

The final step in evaluating $\bar{V}_{i}[\psi]$
is to calculate $\delta H/ \delta \psi_{j}$.
Expressing the free energy
density $F(\rho,{\bf P},{\bf n})$
as a sum of kinetic and potential energies, we rewrite (2.3) as:
\beq
H = \int d^{3} x (\varepsilon_{k}+
\varepsilon_{u}^{\rho}+\varepsilon_{u}^{n}+\varepsilon_{u}^{c})
\eeq
where $\varepsilon_{k}$ is the kinetic energy density of the molecules,
$\varepsilon_{u}^{\rho}$
is the potential energy  density due
to density fluctuations and $\varepsilon_{u}^{n}$
is that due to the director fluctuations.
Finally $\varepsilon_{u}^{c}$ is an energy density due to the coupling of
density and director fluctuations.
For the kinetic energy density we have:
\beq
%\label{3.9}
\varepsilon_{k}({\bf x},t)=\frac{{\bf P}^2({\bf x},t)}{2\rho({\bf x},t
 )},
\eeq
For the potential energy density of the density fluctuations we choose
the simplest form incorporating $\vec{\bigtriangledown}\rho$:
\beq
\varepsilon^\rho_u (\vec{x}, t) ={A\over 2} (\delta\rho(\vec{x},t))^2
+{B\over 2}(\vec{\nabla}\rho (\vec{x}, t))^2
\eeq
where A and B are phenomenological constants, and
$\delta\rho=\rho(\vec{x}, t)-\rho_0$ with $\rho_0$ being the uniform
density. The gradient term in (2.14) is
rotationally isotropic which
is unrealistic in an anisotropic system like the nematic. We will
incorporate the effects of anisotropy below in $\varepsilon^c_u$, the
coupling of $\vec{\nabla}\rho$ to $\hat{n}$

The
simplest choice for $\varepsilon_{u}^{n}$ is:
\beq
\varepsilon_{u}^{n}({\bf x},t) = \frac{1}{2} K ( \nabla_i n_j({\bf x},t))
                                              ( \nabla_i n_j({\bf x},t))
\eeq
where $K$ is a Frank elastic constant and repeated indices are summed
over. In general, the symmetry of a nematic
 allows
for three independent elastic constants$^{11}$ corresponding to the distortions
 splay, twist and
bend. We have made the simplification of setting the three elastic constants
equal to $K$. While this equality is broken upon renormalization (see
Section III), the difference between the elastic constants is not
large and we will ignore it.
\par
Finally in $\varepsilon^c_u$ we incorporate the expected anisotropy in
density fluctuations and choose,
\beq
\varepsilon_{u}^{c}({\bf x},t) = \frac{1}{2} I
({\bf n}({\bf x},t) \cdot {\bf \nabla}\rho({\bf x},t))^{2},
\eeq
where $I$ is a phenomenological coupling constant. In general one might
introduce a coupling of the form
${I\over 2}(\vec{n}\cdot\vec{\nabla}\rho)^2 +
{I^\prime\over 2}(\vec{n}\times\vec{\nabla}\rho)^2$ to account for
the expected anisotropy in the density fluctuations, i.e.\ the energy
of such fluctuations should depend on the relative orientation of the
director and the wavevector of the fluctuation. However, because of
the following vector identity:
\beq(\vec\nabla\rho)^2
=(\vec{n}\cdot\vec{\nabla}\rho)^2 +(\vec{n}\times \vec{\nabla}\rho)^2
\eeq
we can eliminate the coupling proportional to $I^\prime$ in favor of a
redefinition of $I$ and the inclusion of the last term in (2.14). This
choice simplifies our perturbation theory in section III, and allows
$I$ to be negative. However rodlike molecules will probably be
characterized by positive values of $I$ since the director will prefer
to align perpendicular to the wavevector of the density field.
We can think of (2.14) and (2.16) together as providing a simple model
for the static structure factor of the nematic, where density
correlations are not isotropic in space but depend on the local
director orientation. The effect of local structure on a simple fluid
as it is supercooled was considered by Das.$^{18}$
\par
Using (2.9) - (2.17) we can evaluate the reversible terms in (2.1).
For the $\vec{P}$ equation of motion we can express the result
conveniently in terms of the divergence of a reactive stress tensor,
i.e.
\beq
\bar{V}_{\vec{P}i} =-\sum_j \nabla_j\sigma^{P,R}_{ij}, \quad
i,j =x, y, z
\eeq
where
\beqa
\sigma^{P, R}_{ij}  = & {P_i P_j\over \rho}+\delta_{ij}\left[{1\over
2}\chi^{-1}(\delta\rho)^2 +{1\over 2}K(\nabla_k n_\ell)^2\right. \nonumber\\
+ &\left. \rho I\nabla_k (n_k(\vec{n}\cdot\vec{\nabla}\rho))+{I\over
2}(\vec{n}\cdot\vec{\nabla}\rho)^2\right]\nonumber\\
 + & I(\vec{n}\cdot\vec{\nabla}\rho)n_j \nabla_i\rho +K(\nabla_i
n_\ell) (\nabla_j n_\ell)\nonumber\\
 - &  \ \sum_{k} \mu_{ijk} (-K\nabla^2 n_k +J
(\vec{n}\cdot\vec{\nabla}\rho) \nabla_k\rho)
\eeqa
and
\beq
\mu_{ijk} ={1\over 2}(\lambda +1)n_j\delta_{ik}+{1\over
2}(\lambda-1)n_i\delta_{jk}-\lambda n_i n_j n_k
\eeq
The stress tensor appearing in (2.18) is not symmetric. However, the
equation of motion for $\vec{P}$ is not sensitive to this asymmetry. A
symmetric choice for $\sigma^R_{ij}$ would yield the same forces as
well as guaranteeing conservation of angular momentum$^{12}$
 which is not an issue in our analysis.
\par
The reversible part of the director equation of motion is given by,
\beq
\bar{V}_{n_{i}}=-{1\over \rho}\vec{P}\cdot\vec{\nabla}n_i +(\vec{\Omega}\times
\vec{n})_i +\lambda\sum_{k=x, y, z} (n_k\delta_{ij}-n_j n_j
n_k)A_{kj} \quad i=x, y
\eeq
where
\beq
A_{kj} ={1\over 2\rho}(\nabla_k P_j +\nabla_j P_k)
\eeq
and,
\beq
\vec{\Omega}={1\over 2\rho} (\vec{\nabla}\times \vec{P})
\eeq

A dissipative stress tensor can be introduced to write the dissipative
contribution to (2.1). The dissipative
momentum stress tensor $\sigma^{P,D}_{ij}$ is defined via the
relation,
\beq
\sum_{j=x,y,z,} \nabla_j \sigma^{P,D}_{ij}=\sum_{j=x,y,z}
\Gamma_{PiPj} {\delta H\over \delta P_j} \quad i=x,y,z
\eeq
The uniaxial symmetry of the nematic dictates that $\sigma^{P,D}_{ij}$
has the following form,$^{13}$
\beqa
-\sigma_{ij}^{P,D} & = &
  2(\nu^0_1+\nu^0_2-2\nu^0_3) n_k n_m A_{mk} n_i n_j +
2(\nu^0_3-\nu^0_2)(n_i n_k A_{kj}+n_l n_k A_{ik})  \nonumber\\
&&+(\nu^0_5-\nu^0_4+\nu^0_2) (n_i n_j A_{kk}+\delta_{ij}n_k n_m A_{km})+
2 \nu^0_2 A_{ij} +(\nu^0_4-\nu^0_2) A_{kk} \delta_{ij}. \quad\quad
\eeqa
The five bare viscosity coefficients $\nu^0_1, \nu^0_2, \nu^0_3,
\nu^0_4,$ and $\nu^0_5$ are in general not equal to each other. If
$\nu^0_1=\nu^0_2=\nu^0_3=\nu^0_4-\nu^0_5$ we recover the usual form
for the dissipative stress tensor of a simple fluid. If the nematic is
incompressible then all terms in $\sigma^{P,D}_{ij}$ which are
proportional to $A_{kk}$ mush vanish, implying that $\nu^0_2 =\nu^0_4$
and $\nu^0_5=0$, and three independent viscosities remain. As we are
interested in density fluctuations we will work with the full tensor
displayed in (2.25). Comparing (2.24) and (2.25) we can identify the
nonzero elements of the viscosity matrix $\Gamma_{PiPj}$:

\beq
\Gamma_{P_{x} P_{x}} =
-(\nu^0_{4} + \nu^0_{2})\nabla_{x}^{2}- \nu_{3} \nabla_{z}^{2},
\eeq
\beq
\Gamma_{P_{y} P_{y} }   = -\nu^0_{2}\nabla_{x}^{2}-\nu^0_{3} \nabla_{z}^{2},
\eeq
\beq
\Gamma_{P_{z} P_{z}}   =
 -\nu^0_{3}\nabla_{x}^{2}-
(2\nu^0_{1} +\nu^0_{2} -\nu^0_{4}+2\nu^0_{5})\nabla_{z}^{2},
\eeq
\beq
\Gamma_{P_{x} P_{z}} =
    - ( \nu^0_{5} + \nu^0_{3} )\nabla_{x} \nabla_{z}.
\eeq

All other elements of the viscosity matrix $\Gamma_{PiPj}=0.$ If we go
the  limit of a simple fluid $(\nu^0_1=\nu^0_2=\nu^0_3=\nu^0_4-
\nu^0_5)$, then $\Gamma_{PiPj}$ reduces to the tensor $L_{ij}$, used
in refs.\ 5 and 18.
\par
The dissipative contribution to the director equation is proportional
to the   ``molecular field", $\delta H/\delta n_i$, with the
proportionality constant conventionally written as $1/\gamma^0_1$, where
$\gamma^0_1$ has the units of viscosity.  Thus we identify,
\beq
\Gamma_{n_{i}n_{j}}={1\over \gamma^0_1}\delta_{ij}, \quad i,j=x,y
\eeq

All other elements of the viscosity matrix $\Gamma_{ij}$ are zero.
In particular all off-diagonal elements $\Gamma_{n_{i}P_{j}}$ can be
shown  to be zero on the basis of time-reversal symmetry.$^{12}$
\par
The compressible nematic is thus characterized by six hydrodynamic
fields and six independent viscosities: $\nu^0_1, \nu^0_2, \nu^0_3,
\nu^0_4, \nu^0_5$ and $\gamma^0_1$. Another viscosity $\gamma_2$
commonly discussed in the literature which describes the torque
exerted on the director by a shear flow is related to $\gamma_1$ via
$\gamma_2=-\lambda\gamma^0_1$ where $\lambda$ is the reactive
coefficient introduced in eqn. (2.11).
\par
Our final nonlinear hydrodynamic equations for the compressible
nematic are as follows:
\beq
{\partial\rho\over \partial t}+\vec{\nabla}\cdot\vec{P}=0
\eeq
\beq
{\partial P_i\over \partial t}=-\sum_{j=x,y,z}\nabla_j
\sigma^P_{ij}, \quad i=x,y,z
\eeq
\beqa
{\partial n_i\over \partial t}&= & \vec{\Omega}\times \vec{n})_i -{1\over
\rho}(\vec{P}\cdot\vec{\nabla})n_i \nonumber\\
&& +\lambda\sum_{k, \ell=x,y,z} A_{k\ell} n_k (\delta_{i\ell} -n_i
n_\ell)\nonumber\\
&& +{K\over \gamma^0_1} \nabla^2 n_i-{I\over
\gamma^0_1}(\vec{n}\cdot\vec{\nabla}\rho)\nabla_i \rho, \quad i=x,y
\eeqa
where
\beq
\sigma^P_{ij} =\sigma^{P,R}_{ij} +\sigma^{P,D}_{ij}
\eeq
with $\sigma^{P,R}_{ij}$ and $\sigma^{P,D}_{ij}$ given by (2.19) and
(2.25) respectively.

\setcounter{equation}{0}
\section{Transport coefficients}
To investigate the effects of the nonlinearities in the equations of
motion (2.31) -- (2.34) on the transport properties of the nematic we
use the Martin-Siggia-Rose (MSR) formalism.$^{19}$
 This formalism allows us
to calculate the correlation functions
 $G_{ij}(\vec{x}, t;
\vec{x}^\prime, t^\prime) \equiv <\delta\psi_i (\vec{x}, t)
\delta\psi_j (\vec{x}^\prime, t)>$ where $\psi_i$ represents any of
our six hydrodynamic fields and the brackets refer to the average over
the noise source $\Theta_i$ defined in (2.5). It will also enable us the
calculate response functions, and obtain corrections to the bare
viscosities introduced in the previous section. We refer the reader to
refs. 6 and 19 for full details on the MSR method; we summarize the essential
ideas here.
\par
We define a generating function $Z_U[\psi, \hat{\psi}]$ as follows:
\beq
Z_U[\psi ,\hat{\psi}]=C \int D(\psi) D(\hat{\psi}) \;
\exp(-A_U[\psi,\hat{\psi} ])
\; \exp\int d{\bf x} \;dt \; U_i({\bf x},t) \psi_i ({\bf x},t).
\eeq
where C is a constant and $\psi$ collectively represents the six
hydrodynamic fields $\psi_i$. Functional differentiation of $Z_U$ with
respect to U generates the correlation functions
$G_{ij}$:

\beq
G_{ij}(\vec{x}, t; \vec{x}^\prime, t^\prime)= {\delta^2\ell nZ_U\over
\delta U_i(\vec{x}, t)\delta U_j (\vec{x}^\prime, t^\prime)}
\eeq
The six auxiliary fields $\hat{\psi}$ were introduced to exponentiate
each of the six hydrodynamic equations of motion (2.1). The
integration over the noise source $\Theta_i$ has been replaced by an
integration over the fields $\psi_i$. The action $A_U$ is given by:
\beqa
 A_{U}  [\psi,\hat{\psi}] & =&
 \int d\vec{x}d{\bf t}d\vec{x}'dt \hat{\psi}_i({\bf x,t})\beta^{-1}
\Gamma_{ij}({\bf \vec{x}}, t;\vec{x}^\prime, t^\prime)
\hat{\psi}_j({\bf x}';t)\nonumber\\
&&+ i\int d{\bf x}dt \hat{\psi}_j({\bf x},t)
[\frac{\partial \psi_j({\bf x})}{\partial t}
- \bar{H}_j[\psi]].
\eeqa

where $\bar{H}_j$ is defined by:
\beq
 \bar{H}_{j}[\psi] = \bar{V}_j[\psi]-\sum_j\int d{\bf
x}'dt'\Gamma_{ij}({\bf x}'t')
           \frac {\delta H} {\delta \psi_j({\bf x'},t^\prime)}
\eeq
Introducing the "vector"
$\phi= (
      \rho,     {\bf P},      n _{x},     n _{y},
 \hat{\rho},\hat{{\bf P}},\hat{n}_{x},\hat{n}_{y} )$,
we can then rewrite $ A_{U} [\psi, \hat{\psi} ] \equiv A_U [\phi]$ as
follows:
\beqa
%\label{5.7}
A_U [\phi] & = &
\frac{1}{2} \int d1 d2 \sum_{\alpha \beta} \phi_\alpha (1) G_0^{-1}
(1,2) \phi_\beta(2)  \\
&   & + \sum_{N=3...}
\frac{1}{N} \int d1 d2 ...dN \sum_{\underbrace{\alpha \beta \gamma \ldots}_N }
V^{N}_{\underbrace{\alpha \beta \gamma \ldots}_N}
\underbrace {\phi_\alpha (1) \phi_\beta(2) \phi_\gamma(3)\ldots}_N \nonumber.
\eeqa
Here the integration variables $1,2,3,\ldots$ stand for $({\bf x},t)$.
The explicit expressions $G^{-1}_0$ for the vertices $V_{\alpha \beta
\gamma...} $
will be presented later.  The action $A_U$ generates the full non-linear
hydrodynamic equations for a nematic liquid. When we omit the vertices what
remains then is a pure quadratic Gaussian field theory.
The corresponding $A_U$ contains the inverse of the linearized
correlation matrix $G_0^{-1}$ and thus generates the linearized hydrodynamic
equations of a nematic liquid.   We first discuss this limit before
doing perturbation theory in the vertices.
\par
The linearized theory, as well as the perturbation theory, are most
easily discussed in Fourier space. We Fourier transform (3.5) in space
and time recalling our discussion following (2.5) where we assumed
that the wavevector $\vec{k}$ lies in the $x-y$ plane. The matrix
$G^{-1}_0 (\vec{k}\omega)$ is displayed in Table I  where we have
introduced the following tensors for  notational convenience:
\beqa
\alpha_{ij} &= & \delta_{ij}{K\over 2}(\lambda +1)k_z\nonumber\\
&& + \delta_{ix}\delta_{jz}{K\over 2}(\lambda -1)k_x
\eeqa
\beqa
L^0_{ij}&=&\delta_{ij}\delta_{jx}\Gamma^0_{xx}+\delta_{iy}\delta_{jy}
\Gamma^0_{yy}\nonumber\\
&+&\delta_{iz}\delta_{jz}\Gamma^0_{zz}+(\delta_{ix}\delta_{jz}+
\delta_{ij}\delta_{jz})\Gamma^0_{xz}
\eeqa
where $\Gamma^0_{xx}, \Gamma^0_{yy}, \Gamma^0_{zz}$ and
$\Gamma^0_{xz}$ are the Fourier transforms of the viscosity matrix
elements (2.26) -- (2.29) respectively.
\par
The inversion of $G^{-1}_0$ is most easily accomplished by decomposing
$\vec{P}$ and $\vec{n}$ into longitudinal and transverse components.
The calculation is straightforward though tedious especially for the
longitudinal portion. The elements of the matrix $G_0$ provide the
physical correlation and response functions for the linearized theory.
Correlation functions of any two hatted variables are identically zero.
Correlation functions of unhatted variables can be found to leading
order in $k^2$ from the corresponding response functions via the
fluctuation-dissipation theorems:
\beq
<\rho\rho>=-2\beta^{-1}\chi Im<\rho\hat{\rho}>
\eeq
\beq
<n_x n_x> =-2\beta^{-1}\chi_n Im<n_x \hat{n}_x>
\eeq
\beq
<n_y n_y>=-2\beta^{-1}\chi_n Im<n_y \hat{n}_y>
\eeq
where $\chi$ is the static density structure factor given by
$(A+Bq^2)^{-1}$ and $\chi_n$ is the static director structure factor
given by $(Kk^2)^{-1}$. We could also write corresponding relations
for the momentum correlation functions, however, they are not needed
in elucidating the mode structure and calculating the corrections to
the transport coefficients.
\par
The density and director response functions in the linearized theory
are given by the following expressions:
\beq
<\delta\rho (\vec{k},\omega)\delta\hat{\rho}(-\vec{k},-\omega)>^o=
{\omega\rho_o+ik^2\Gamma_o\over \rho_o(\omega^2-
c^2_ok^2)+i\omega k^2\Gamma_o}
\eeq
\beq
<\delta n_x(\vec{k}\omega)\delta\hat{n}_x(-\vec{k},-\omega)>^o=
{\omega\rho_o+ik^2\nu_L^o\over \rho_o(\omega+ik^2\tilde{\Gamma}^o_s)
(\omega+ik^2\tilde{\Gamma}^o_f)}
\eeq
\beq
<\delta n_y(\vec{k},\omega)\delta n_y(-\vec{k}, -\omega)>^o=
{\omega\rho_o+ik^2\nu_T^o\over \rho_o(\omega+ik^2\Gamma^o_s)
(\omega+ik^2\Gamma^o_f)}
\eeq
The results are correct up to terms of relative order $k^2$. The
viscosities  appearing in (3.11)--(3.13) are given by:
\beqa
\Gamma_o &=& -(\nu^o_1+\nu^o_2 -2\nu^o_3)\sin^2(2\theta)/2\nonumber\\
&&+(\nu^o_2+\nu^o_4)\sin^2\theta\nonumber\\
&&+(2\nu^o_1+2\nu^o_5 +\nu^o_2 -\nu^o_4)\cos^2\theta
\eeqa
\beq
\nu^o_L=\nu^o_3\cos^2(2\theta)+(\nu^o_1+\nu^o_2)\sin^2(2\theta)/2
\eeq
\beq
\nu^o_T=\nu^o_2\sin^2\theta+\nu^o_3\cos^2\theta
\eeq
\beqa
\tilde{\Gamma}^o_{s,f} &=& {1\over 2}\left({\nu^o_L\over \rho_o}+{K\over
\gamma_1}\right)\pm \nonumber \\
& &{1\over 2}\sqrt{\left({\nu^o_L\over \rho_o}+{K\over
\gamma_1}\right)^2 -\left( (1+\lambda\cos 2\theta)^2{K\over \rho_o}+
{4K\nu^o_L \over \rho_o\gamma_1}\right)}
\eeqa
\beqa
\Gamma^o_{s,f}&=&{1\over 2}\left({\tilde{\Gamma}_b\over \rho_o} +{K\over
\gamma_1}\right)\pm\nonumber\\
&&{1\over 2}\sqrt{\left({\tilde{\Gamma}_b\over \rho_o}+{K\over
\gamma_1}\right)^2 -\left( (\lambda+1)^2{K\over \rho_o}\cos^2\theta +
{4K\tilde{\Gamma}_b\over \rho_o\gamma_1}\right)}
\eeqa
where $\theta$ is the angle between $\vec{k}$ and the z axis.
\par
In writing (3.17) and (3.18) we have assumed that the director modes
appearing in (3.12) and (3.13) are diffusive rather than propagating.
This is true for equilibrated nematics$^{13}$ where the orientational
relaxation time of the director $(Kk^2/\gamma_1)^{-1}$ is small
compared to the shear diffusion times $(\nu^o_L k^2/\rho_o)^{-1}$ and
($\nu^o_T k^2/\rho_o)^{-1}$. We shall see subsequently that this does
not remain true when the nematic is supercooled or compressed rapidly,
and propagating  shear modes can appear. In the absence of
supercooling (3.17) and (3.18) can be approximated as follows:
\beq
\tilde{\Gamma}^o_f \approx \nu^o_L/\rho_o
\eeq
\beq
\tilde{\Gamma}^o_s \approx \Gamma^o_s \approx K/\gamma_1
\eeq
\beq
\Gamma^o_f \approx \nu^o_T/\rho_o
\eeq
Thus there are two ``slow" modes with viscosities  $\Gamma^o_s$ and
$\tilde{\Gamma}^o_s$ corresponding to the slow relaxation of director
fluctuations, while the fast modes with viscosities  $\Gamma_f$ and
$\tilde{\Gamma}_f$ are like ordinary shear waves. In addition to these
four modes we also have two sound modes appearing in (3.11) with speed
$c_o$ and damping $\Gamma_o$.
\par
We now calculate the corrections to the linearized theory due to the
vertius $V^N$ appearing in (3.5). If we perform the rescalings
$\psi\rightarrow\beta^{+1/2}\psi$ and $\hat{\psi}\rightarrow\beta^{-
1/2}\hat{\psi}$ we see that  the quadratic part of $A_U$ is
$0((k_BT)^o)$, and the higher-order terms proportional to $V^N$ are of
order $(k_B T)^{(N/2)-1}$. Thus we can systematically compute
corrections to the linearized theory in powers of $k_BT.$ In
particular one-loop diagrams will be $0(k_B T)$.
\par
The inverse of the correlation matrix for the complete nonlinear
theory satisfies the formal equation:
\beq
G^{-1}(1,2)=G_o^{-1}(1,2)-\Sigma(1,2)
\eeq
which defines the self-energy $\Sigma$. We can then write
corresponding equations for the renormalized transport coefficients by
referring to Table I. As shown in ref.\ 6 the renormalized viscosities
are most readily obtained by looking at the renormalization of the
terms  in the action which  are quadratic in the hatted fields. Thus
the elements of the viscosity matrix $\Gamma_{ij}$ renormalize as
follows:
\beq
\Gamma^{\vec{k},\omega}_{xx}=\Gamma^o_{xx}+{\beta\over
2}\Sigma_{\hat{P}_x\hat{P}_x} (\vec{k},\omega)
\eeq
\beq
\Gamma_{yy}(\vec{k},\omega)=\Gamma^o_{yy}+{\beta\over
2}\Sigma_{\hat{P}_y\hat{P}_y}(\vec{k},\omega)
\eeq
\beq
\Gamma_{zz}(\vec{k},\omega)=\Gamma^o_{zz}+{\beta\over
2}\Sigma_{\hat{P}_z\hat{P}_z}(\vec{k},\omega)
\eeq
\beq
\Gamma_{xz}(\vec{k},\omega)=\Gamma^o_{xz}+{\beta\over
2}\Sigma_{\hat{P}_x\hat{P}_z}(\vec{k},\omega)
\eeq
\beq
1/\gamma_1 (\vec{k},\omega)={1\over \gamma^o_1}+{\beta\over
2}\Sigma_{\hat{n}_x\hat{n}_x}(\vec{k}\omega)={1\over
\gamma^o_1}+{\beta\over 2}\Sigma_{\hat{n}_y\hat{n}_y}(\vec{k},\omega)
\eeq

Referring to Table I we find that $K/\gamma_1, \lambda$ and $C$
renormalize as follows:
\beqa
{K\over \gamma_1} & =& \left({K\over \gamma_1}_o\right)+{i\over
k^2}\Sigma_{n_x\hat{n}_x}(\vec{k},\omega)\nonumber\\
&=&\left({K\over \gamma_1}\right)_o +{i\over k^2}\Sigma_{n_y
\hat{n}_y}(\vec{k},\omega
\eeqa
\beq
\lambda=\lambda_o -{2\over k_2}\Sigma_{\hat{n}_y P_y}(\vec{k},
\omega)
\eeq
\beq
c^2=c^2_o=+{1\over k_z}\Sigma_{\hat{P}_x\rho}
\eeq
The uniaxial symmetry of the nematic as well as conservation of
momentum allows us to  the write momenta self-energies as follows, (cF.
(2.26)--(2.29)):
\beqa
\Sigma_{\hat{P}_x\hat{P}_x}(\vec{k,\omega})= &-&k^2_x
(\gamma_2(\vec{k},\omega)+\gamma_4(\vec{k},\omega))\nonumber\\
&-&k^2_z\gamma_3(\vec{k},\omega)
\eeqa
\beq
\Sigma_{\hat{P}_y\hat{P}_y}(\vec{k},\omega)=-k^2_x\gamma_2
(\vec{k},\omega)-k^2_z\gamma_3(\vec{k},\omega)
\eeq
\beqa
\Sigma_{\hat{P}_z\hat{P}_z}(\vec{k},\omega)&=& -
k^2_x\gamma_3(\vec{k},\omega)\nonumber\\
&& -k^2_z\left[2\gamma_1
(\vec{k},\omega)+\gamma_2(\vec{k},\omega)\right.\nonumber\\
&&\left. -\gamma_4(\vec{k},\omega)+2\gamma_5 (\vec{k},\omega)\right]
\eeqa
\beq
\Sigma_{\hat{P}_x\hat{P}_z}(\vec{k},\omega)=-
k_zk_z(\gamma_3(\vec{k},\omega)+\gamma_5(\vec{k},\omega))
\eeq
where the functions $\gamma_i (\vec{k}, \omega), i=1,\ldots, 5$
renormalize the viscosities $\nu_i$ as follows:
\beq
\nu_i (\vec{k}, \omega)=\nu^o_i +{\beta\over 2} \gamma_i(\vec{k},
\omega), i=1, \ldots, 5
\eeq
\par
In the hydrodynamic limit the response functions will have the same
form as in the linearized theory with the bare transport coefficients
replaced by their renormalized values which depend on $\vec{k}$ and
$\omega$. As discussed in the Introduction we are ignoring any
possible nonhydrodynamic terms which might cutoff a sharp transition
and focus instead  on the behavior of the renormalized viscosities.
Eqns. (3.14) -- (3.18) will then be valid for the renormalized
quantities and we have the following relations for the generalized
transport coefficients appearing in the response functions:
\beqa
\Gamma(\vec{k}, \omega)=\Gamma^o-&\beta&\left[{k^2_x\over
k^4}\Sigma_{\hat{P}_x\hat{P}_x}(\vec{k},\omega)+{k^2_z\over
k^4}\Sigma_{\hat{P}_z\hat{P}_z}(\vec{k}, \omega)\right.\nonumber\\
&&\left.+2{k_x k_z\over
k^4}\Sigma_{\hat{P}_x\hat{P}_z}(\vec{k},\omega)\right]
\eeqa
\beqa
\nu_L(\vec{k},\omega)=\nu^o_L- &\beta& \left[{k^2_z\over
k^4}\Sigma_{\hat{P}_x\hat{P}_x} (\vec{k},\omega)\right.\nonumber\\
&&\left.+{k^2_x\over k^4}\Sigma_{\hat{P}_z \hat{P}_z}(\vec{k}, \omega)-{2k_x
k_z\over k^4}\Sigma_{\hat{P}_x \hat{P}_z}(\vec{k}, \omega)\right]
\eeqa
\beq
\nu_T (\vec{k},\omega) =\nu^o_T -\beta\left({1\over
k^2}\right)\Sigma_{\hat{P}_y\hat{P}_y}(\vec{k},\omega)
\eeq
Using (3.32) -- (3.35) we see that these renormalized viscosities are
well behaved in the limit $\vec{k}\rightarrow 0$.
\par
The preceding discussion indicates that we need to calculate the
following self-energies in order to obtain the renormalized transport
coefficients:
$\Sigma_{\hat{P}_x\hat{P}_x},\Sigma_{\hat{P}_y\hat{P}_y},
\Sigma_{\hat{P}_z\hat{P}_z},\Sigma_{\hat{P}_x\hat{P}_z},
\Sigma_{\hat{n}_x\hat{n}_x}, \Sigma_{n_{x}\hat{n}_{x}},
\Sigma_{\hat{n}_y P_y},$ and
$\Sigma_{\hat{P}_x \rho}$. We have done so to one-loop order under the
following condition: we keep only those diagrams where there is a
possibility of feedback from either density or director fluctuations.
Thus we consider diagrams where the propagators are either
$G_{\rho\rho}(\vec{k},\omega)$ or $G_{n_{i}n_{j}} (\vec{k},\omega), i,
j=x,y$. (In the Appendix  we show that mixed propagators cannot yield a
feedback). The diagrams will be bubble-type and contain two
three-point vertices. The symmetrized three-point vertices in the
action (3.5) that contribute to the diagrams within our approximation
are given by:
\beqa
V_{\alpha\beta\gamma}&=&{1\over 2}\left[\tilde{V}_{\alpha\beta\gamma}
(1,2,3)+ \tilde{V}_{\beta\alpha\gamma}(2,1,3)\right.\nonumber\\
&&+\tilde{V}_{\gamma\beta\alpha}(3,2,1)+
\tilde{V}_{\alpha\gamma\beta}(1,3,2)\nonumber\\
&&\left.+\tilde{V}_{\beta\gamma\alpha}(2,3,1)+
\tilde{V}_{\gamma\alpha\beta}(3,1,2,)\right]
\eeqa
where
\beq
\tilde{V}_{\alpha\beta\gamma}(1,2,3)=\sum_{i=1}^{11}
\tilde{V}_{\alpha\beta\gamma}^{(i)} (1,2,3)
\eeq
and the $\tilde{V}^{(i)}$ are given in Fourier space by,

\beqa
%\label{5.11}
\tilde {V}^{(1)}_{\alpha, \beta,\gamma}(1,2,3) &=&  \frac{i}{2}
\sum_{ij} \delta _{\alpha \hat{P_i}} \delta _{\beta \rho}\delta _{\gamma \rho}
i\cdot \nonumber\\
&&[A\delta_{ij}(k_{2}+k_{3})_j +B((k_2 +k_3)_j
k_{3i}k_{3j}\nonumber\\
&&-(k_2 +k_3)_i k_{2j}k_{3j})]\delta(1+2+3) (2\pi)^4
\eeqa
\beq
\tilde {V}^{(2)}_{\alpha, \beta,\gamma}(1,2,3) = i K
\sum_{jl} \delta _{\alpha \hat{p_j}} \delta _{\beta n_l}\delta _{\gamma n_l}
(i)^3 k_{2j}k^2_{3}\delta(1+2+3) (2\pi)^4
\eeq
\beq
\tilde {V}^{(3)}_{\alpha, \beta,\gamma}(1,2,3) = \frac{i}{2}(\lambda +1) K
\sum_{jl} \delta _{\alpha \hat{p_j}} \delta _{\beta n_l}\delta _{\gamma n_j}
(i)^3 (k_{2l} +k_{3l})k^2_{3}\delta(1+2+3) (2\pi)^4
\eeq
\beq
\tilde {V}^{(4)}_{\alpha, \beta,\gamma}(1,2,3) = \frac{i}{2} (\lambda -1) K
\sum_{jl} \delta _{\alpha \hat{p_j}} \delta _{\beta n_j}\delta _{\gamma n_l}
(i)^3 (k_{2l}+k_{3l})k^2_{3}\delta(1+2+3) (2\pi)^4
\eeq
\beq
\tilde {V}^{(5)}_{\alpha, \beta,\gamma}(1,2,3) = -i \lambda K
\sum_j \delta _{\alpha \hat{p_z}} \delta _{\beta n_j}\delta _{\gamma n_j}
(i)^3 (k_{2z}+k_{3z})k^2_{3}\delta(1+2+3) (2\pi)^4
\eeq
\beq
\tilde {V}^{(6)}_{\alpha, \beta,\gamma}(1,2,3) = -i I
\sum_j \delta _{\alpha \hat{p_j}} \delta _{\beta \rho}\delta _{\gamma\rho}
(i)^3 (k_{3j}k^2_{3z})\delta(1+2+3) (2\pi)^4
\eeq
\beq
\tilde {V}^{(7)}_{\alpha, \beta,\gamma}(1,2,3) = -i \frac{I}{2} (\lambda +1)
\sum_j\delta _{\alpha \hat{p_j}} \delta _{\beta \rho}\delta _{\gamma\rho}
(i)^3 (k_{2j}k_{3z})(k_{2z}+k_{3z})\delta(1+2+3) (2\pi)^4
\eeq
\beq
\tilde {V}^{(8)}_{\alpha, \beta,\gamma}(1,2,3) = -i \frac{I}{2} (\lambda -1)
\sum_j\delta _{\alpha \hat{p_z}} \delta _{\beta \rho}\delta _{\gamma\rho}
(i)^3 (k_{2j}k_{3z})(k_{2j}+k_{3j})\delta(1+2+3) (2\pi)^4
\eeq
\beqa
\tilde {V}^{(9)}_{\alpha, \beta,\gamma}(1,2,3) = -i I \rho^0
\sum_{j}\delta _{\alpha \hat{p_j}} \delta _{\beta n_x}\delta_{\gamma\rho}
(i)^3 (k_{2j}+k_{3j})\cdot ((k_{2x}+k_{3x})k_{3z}+\nonumber\\
(k_{2z}+k_{3z})k_{3x})\delta(1+2+3)
(2\pi)^4
\eeqa
\beq
\tilde {V}^{(10)}_{\alpha, \beta,\gamma}(1,2,3) = i \lambda I
\delta _{\alpha \hat{p_z}} \delta _{\beta \rho}\delta _{\gamma\rho}
(i)^3 (k_{2z}+k_{3z})k_{2z}k_{3z}\delta(1+2+3) (2\pi)^4 .
\eeq
\beqa
\tilde{V}^{11}_{\alpha\beta\gamma}(1,2,3) &=& -i{I\over
\gamma_1}\sum_{\ell}\delta_{\alpha\hat{n}_\ell}\delta_{\beta\rho}
\delta_{\gamma\rho}(i)^2\nonumber\\
&&k_{2\ell}k_{3z}\delta(1+2+3)(2\pi)^4
\eeqa
where $\delta(1+2+3)\equiv \delta(\vec{k}_1+\vec{k}_2+\vec{k}_3)$. The
three-point vertices appearing in (3.42) -- (3.51) arise from the
reactive momentum stress tensor (2.19) excluding the convective term
$(P_iP_j/\rho)$. The vertex (3.52) arises from the dissipative part of
the director equation; as we shall see it ultimately plays no role in
the growth of the  viscosities.
\par
Using these vertices we have evaluated the relevant self-energies. The
results are quite complex and are tabulated in the Appendix. The
renormalized transport coefficients have the following form:
\beqa
\Gamma(\vec{k},\omega)&=&\Gamma_o+{\beta\over 2}\int^\infty_0
dt \ e^{iwt}\int{d\vec{k}^\prime\over (2\pi)^3}\nonumber\\
&&\times\left[G_{\rho\rho}(\vec{k}^\prime,t)G_{\rho\rho}(\vec{k}-
\vec{k}^\prime,t)(\chi^{-
2}+f_o(\vec{k},\vec{k}^\prime)\right]\nonumber\\
&&+G_{n_xn_x}(\vec{k}^\prime,t)G_{n_x n_x}(\vec{k}-
\vec{k}^\prime,t)K^2f_1(\vec{k}\vec{k}^\prime)\nonumber\\
&&+G_{n_yn_y}(\vec{k}^\prime,t)G_{n_yn_y}(\vec{k}-\vec{k}^\prime
t)K^2f_2(\vec{k},\vec{k}^\prime)\nonumber\\
&& \left.+G_{\rho\rho}(\vec{k}^\prime,t)G_{n_x n_x}(\vec{k}-
\vec{k}^\prime,t)I^2f_3(\vec{k},\vec{k}^\prime)\right]
\eeqa
\beqa
\nu_L(\vec{k},\omega)&=&\nu^o_L+{\beta\over 2} \int^\infty_o dt \
e^{i\omega t}\int{d\vec{k}^\prime\over (2\pi)^3}\nonumber\\
&&\times \left[G_{\rho\rho}(\vec{k}^\prime,t)G_{\rho\rho}(\vec{k}-
\vec{k}^\prime,t) K^2g_1(\vec{k},\vec{k}^\prime)\right.\nonumber\\
&&\left.+G_{n_yn_y}(\vec{k}^\prime,t)G_{n_yn_y}(\vec{k}-
\vec{k}^\prime,t)K^2 g_2(\vec{k},\vec{k}^\prime)\right]
\eeqa
\beqa
\nu_T(\vec{k},\omega)&=&\nu^o_T +{\beta\over 2}K^2k^2_x\int^\infty_o dt
\ e^{i\omega t}\nonumber\\
&&\int{d\vec{k}^\prime\over (2\pi)^3}
\left[h_1(\vec{k},\vec{k}^\prime)G_{n_x
n_x}(\vec{k}^\prime, t)\right.\nonumber\\
&&\left.\times G_{n_yn_y}(\vec{k}-\vec{k}^\prime, t)+h_2
G_{\rho\rho}(\vec{k}^\prime ,t)G_{\rho\rho} (\vec{k}-\vec{k}^\prime
,t)\right]
\eeqa
where $G_{\rho\rho}(\vec{k}, t), \ G_{n_xn_x}(\vec{k}, t), \ G_{n_y
n_y}(\vec{k},t)$ are the correlation functions of the density and
director modes respectively as functions of $\vec{k}$ and $t$. They
can be found by inverse Laplace transforming the response functions.
The momentum-dependent functions $f_o, f_1, f_2, f_3, g_o, g, g_2$
$h_1$and $h_2$ are discussed in the Appendix ; their precise form is
not necessary
here. However, we do note that their angular dependence implies that
all of the viscosities $\nu_i, i=1, \ldots 5$ are subject to density
and director feedback. On the other hand, there is \underbar{no}
feedback for $c, \lambda$, and most importantly $K/\gamma_1$. As we
shall see in the next section, this latter result is important in
eliminating the possibility of director freezing.

\setcounter{equation}{0}
\section{Implications for Supercooling}
We now consider the implications of our results from the previous
section for the supercooling of a nematic liquid crystal. Our analysis
follows Leutheusser's original approach to the density feedback
mechanism, except we are also interested in a potential director
feedback mechanism, and the behavior of the six viscosity
coefficients.
\par
We begin by recalling Leutheusser's argument for the freezing of the
density fluctuations in a simple fluid. The density response function
(3.11) can be rewritten as,
\beqa
\Phi_1(\vec{k},\omega) &\equiv& <\delta\rho(\vec{k},\omega)\delta\hat{\rho}
(-\vec{k},-\vec{\omega})>\nonumber\\
&&={1\over \omega-{\rho k^2 c^2\over \omega\rho+ik^2\Gamma}}
\eeqa
If the viscosity $\Gamma$ grows as the fluid is supercooled (or
compressed), then (4.1) indicates that
\beq
\Phi_1(k,\omega)\sim{1\over \omega+i\rho
c^2/\Gamma},\Gamma\rightarrow\infty
\eeq
This form indicates that the liquid is freezing, in particular as
$\Gamma\rightarrow\infty, \ \Phi_1$ develops a pole at $\omega=0$. In
deriving (4.2) we assumed that $\Gamma$ was growing very large.
Leutheusser showed that this will in fact occur via a feedback
mechanism that couples density fluctuations to $\Gamma$, specifically
the result (3.53), (dropping the terms proportional to $G_{nn}$ for
the moment). The correlation function $G_{\rho\rho}(\vec{k},t)$ is
given by the inverse Laplace transform of $\Phi_1$ multiplied  by
$\chi$. Thus, ignoring the $\vec{k}$ dependence, (3.53) yields on
equation of the form,
\beq
\Gamma(\omega)=\Gamma^o +\lambda\int^\infty_o dt \ e^{i\omega
t}\Phi^2_1 (t)
\eeq
where $\Phi_1(t)$ is the inverse Laplace transform of
$\Phi_1(\omega)$. Equations (4.2) and (4.4) can be solved to yield a
glass transition where $\Phi_1(\omega)\sim 1/\omega, \
G_{\rho\rho}(t)\rightarrow$
nonzero constant, as $t\rightarrow\infty$  and
$\Gamma(\omega=0)$ diverges. In particular $\Gamma(\omega=0)$ diverges
as,
\beq
\Gamma(\omega=0)\sim(T-T_G)^\mu
\eeq
where $\mu\approx 1.8$ and $T_G$ is the glass transition temperature.
\par
As discussed in the Introduction it is now believed that (4.2) is not
correct and should be replaced by,
\beq
\Phi_1 (\vec{k}, \omega)\sim {1\over \omega+i\rho c^2/\Gamma+i\gamma}
\eeq
where $\gamma$ does not go to zero as the fluid is supercooled. The
origin of $\gamma$ is still a subject of debate.$^{6.7}$ Its presence will
cutoff the Leutheusser transition and (4.4) will not be true
asymptotically. Nevertheless, it is believed that there will be a
substantial growth in the viscosity as the fluid is supercooled which
is eventually rounded off.
\par
We now proceed with a Leutheusser style analysis for the nematic,
bearing the preceding discussion in mind about the limits of such an
analysis. We first examine the longitudinal director response function
given in (3.12)

\beqa
\Phi_2(\vec{k},\omega) & \equiv & <\delta n_x(\vec{k},\omega)\delta\vec{n}_x
(-\vec{k},-\vec{\omega}>\nonumber\\
&&={1\over  \omega+{ik^2\omega{K\rho o\over \gamma_1}-k^4 (\nu_L{K\over
\gamma_1} +{1\over 4}(1+\lambda\cos 2\theta)^2 K)\over
\omega\rho_o +ik^2 \nu L}}
\eeqa
where we have used (3.17). Equation (3.54) indicates that $\nu_L$ does
incorporate density feedback (through the product
$G_{\rho\rho}(\vec{k}^\prime, t) G_{\rho\rho}(\vec{k}-
\vec{k}^\prime,t))$ and hence it will grow as the nematic is
supercooled. Eqn. (4.6) reduces then in the large $\nu_L$ limit to:
\beq
\Phi_2(\vec{k}, \omega)\sim{1\over \omega +ik^2 K/\gamma_1}, \quad
\nu_L\rightarrow \infty
\eeq
 In the  Appendix  we show that $K/\gamma_1$ does
\underbar{not} renormalize in any dramatic way and is not affected by
the density or possible director feedback mechanism. Experimentally,
$K/\gamma_1$ has been measured in supercooled nematics$^{20}$ and found to be
consistent with $K/\gamma_1\sim e^{-1/T}$. Thus, as the temperature is
reduced, the director mode will slow down but there will be no sharp
transition a $\ell a$ Leutheusser or even a rounded one. The viscosity
$\nu_L$ will grow and appear to diverge due to the density
fluctuations as will $\nu_T$. The absence of a freezing transition for
the director modes is true even if the director modes are propagating
rather than diffusive. In the limit of large $\nu_L$, (4.7) is still
obtained even if $\tilde{\Gamma}_s$ and $\tilde{\Gamma}_f$ in (3.17)
become complex (which implies that there are damped, propagating
modes). Thus to study the growth of the viscosities in (3.53) --
(3.55), all terms proportional to either $G_{n_x n_x}$ or $G_{n_y
n_y}$ can be dropped, and feedback from density fluctuations alone
occurs.
\par
We now discuss the experimental implications of our results. First we
summarize our results in terms of the viscosity coefficients
$\nu_1, \nu_2,\nu_3,\nu_4,\nu_5, \gamma_1$ and $\gamma_2$. The latter
two viscosities as
we have already indicated show no pretransitional  growth, at least
within the context of our theory. As noted above, the transport
coefficient $K/\gamma_1$ has been measured in supercooled nematics and
only activated behavior is observed. However, we do expect that the
five viscosities $\nu_i$ will show significant growth as the nematic
is supercooled. As they are all driven by the same density feedback
mechanism we expect them to show the \underbar{same} power-law
behavior indicated  in (4.4), with eventual rounding off. At this time
we cannot predict how wide the temperature regime will be where (4.4)
is valid. We also expect on the basis of the results of refs. 18 and
21   where
the effects of local structure on a simple fluid were considered that
the shear modes of the nematic will become propagating modes at
sufficiently high frequency.  If the Leutheusser scenario were not
ultimately invalidated by a cutoff mechanism then at the glass
transition and below the shear modes would propagate at any frequency
and a nonzero shear modulus would be present. Again at this time we cannot
predict what the lower frequency cutoff will be; calculation of this
cutoff will be sensible once the controversy regarding the rounding
off of the glass transition is resolved. At that time it will also be
sensible to study in detail (3.53) -- (3.55) and calculate the
$\vec{k}$ and $\omega$ dependence of the viscosities.
The modes associated with the director
relaxation remain diffusive. In that sense there will be an
interchange of the ``fast" and ``slow" modes of eqns. (3.19) --
(3.21), with director relaxation becoming fast and shear relaxation
becoming slow.

\bigskip
Table I. The zeroth order matrix $G^{-1}_{0}$. The tensors
$\alpha_{ij}$ and $L_{ij}$ are defined in eqns. (3.6) and (3.7), and
$\chi^{-1}=A+Bk^2$

\bigskip
\begin{center}
\begin{tabular}{l|lllrrr}
& $\rho$ & $P_i$ & $n_j$ & $\hat{\rho}$& $\hat{P}_j$ & $\hat{n}_j
$\\\hline
$\rho$& 0& 0& 0& $-\omega$& $\rho_o\chi^{-1}k_j$ & 0\\
$P_i$& 0& 0& 0& $k_i$ &$-\omega\delta_{ij}+iL_{ij}/\rho_o$ & $-{1\over
\rho_o} K\alpha_{ij}$ \\
$n_i$ &0 &0 &0 &0 &$-\alpha_{ij} k^2$ & $\delta_{ij}(\omega
+{iKk^2\over \gamma^1})$\\
$\hat{\rho}$ & $\omega$ & $-k_i$ & 0& 0& 0& 0\\
$\hat{P}_i$ & $-\rho_o\chi^{-1}k_j$&
$\omega\delta_{ij}+iL_{ij}/\rho_o$ & $\alpha_{ji}k^2$ &
0&$2\beta^{-1}L_{ij}$&0\\
$\hat{n}_i$ & 0& ${1\over \rho_o K}\alpha_{ij}$ &$\delta_{ij}(\omega
+{iKk^2\over \gamma_1})$
& 0 &0 &$2\beta^{-1}\gamma_1^{-1}$
\end{tabular}
\end{center}

\vfil\eject
\bigskip
\noindent
{\bf Appendix}

In this appendix we provide the technical details of our one-loop
calculation of the self-energies appearing in (3.24) -- (3.31).
\par
We consider only those diagrams where the propagators are either
$G_{\rho\rho}(\vec{k},\omega)$ or $G_{nn}(\vec{k},\omega)$ in order to
search for a growth in the viscosities due to density and/or director
feedback. We will ignore mixed propagators $G_{\rho n_x}$. As
discussed by Forster, this correlation function contains both sound
poles as well as $\tilde{\Gamma}_s$ and $\tilde{\Gamma}_f$. When
expanded in terms of these poles, the strengths of the sound poles can
be determined and for large viscosities yield a constant contribution,
rather than the 1/$\omega$ needed for feedback. The strengths of the
director poles cannot be determined to the order in $k$ that we have
worked, but it is unlikely that a 1/$\omega$ pole would emerge.
\par
The diagrams potentially yielding feedback are all of the bubble
variety formed from two three-point vertices. In equations (3.42) --
(3.52) we have listed those vertices which contribute to the
self-energies we need. While we have chosen the external momentum
$\vec{k}$ to lie in the $x-y$ plane, we note that internal momenta can
point in any direction.
\par
We find the following results for the self-energies with two hatted
external lines:

\begin{eqnarray*}
\Sigma_{\hat{P}_x \hat{P}_x} & = &\int {d^3 k^\prime\over
(2\pi}^3{d\omega^\prime\over 2\pi}\left\{G_{\rho\rho}(\vec{k}^\prime,
\omega^\prime) G_{\rho\rho}(\vec{k}-\vec{k}^\prime,\omega-
\omega^\prime)\right. \nonumber \\
&& \cdot  [{1\over 4}(Ak_x +B(k_x^\prime(\vec{k}\cdot \vec{k}^\prime)-
k_x\vec{k}^\prime\cdot(\vec{k}-\vec{k}^\prime)))\cdot\nonumber\\
&& (2Ak_xB(k_x^\prime k^{\prime 2}+k_x
\vec{k}^\prime\cdot(\vec{k}^\prime - \vec{k}))+\nonumber\\
&&-I^2(k_x -k_x^\prime)(k_z - k^\prime_z)^2 \left((k_x -k_x^\prime)(k_z -
k_z^\prime)^2 +k_x^\prime k_z^{\prime 2}\right)\nonumber\\
&&-I^2(\lambda+1)\left({1\over 4}(\lambda +1) k_z^2k_x^\prime (k_z -
k_z^\prime)\right.\nonumber\\
&&\left.+k_z(k_x -k_x^\prime)(k_z -k_z^\prime)^2)\cdot(k_x^\prime
(k_z -k_z^\prime)+k_z^\prime (k_x-k_x^\prime)\right)\nonumber\\
&&-\left(I((k_x-k_x^\prime)(k_z -k_z^\prime)^2 +k_x^\prime
k_z^{\prime 2})\right.\nonumber\\
&&\left.+{I\over 2} (\lambda +1)k_z (k_x^\prime(k_z -
k_z^\prime)+k_z^\prime(k_x-k_x^\prime))\right)\nonumber\\
&&\cdot(Ak_x+B(k^\prime_x (\vec{k}\cdot\vec{k}^\prime)-k_x
\vec{k}^\prime \cdot(\vec{k}-\vec{k}^\prime))\nonumber\\
&&+G_{n_x n_x}(\vec{k}^\prime,\omega^\prime)G_{n_x n_x}(\vec{k}-
\vec{k}^\prime,\omega-\omega^\prime)\nonumber\\
&&\cdot\left\{-K^2k_x^\prime (\vec{k}-\vec{k}^\prime)^2(k_x^\prime
(\vec{k}-\vec{k}^\prime)^2 +k^{\prime 2}(k_x -k_x))\right.\nonumber\\
&&\left.-(2\lambda K^2k_x +\lambda^2 K^2 k_x^2)(\vec{k}-\vec{k}^\prime)^2
(k^{\prime 2}+(\vec{k}-\vec{k}^\prime)^2 \right \}\nonumber\\
&&+ G_{n_y n_y}(\vec{k}^\prime,\omega)G_{n_y n_y}(\vec{k}-
\vec{k}^\prime, \omega-\omega^\prime)\nonumber\\
&&\left.\cdot (-K^2 k_x^\prime (\vec{k} -\vec{k}^\prime)^2 (k_x^\prime
(\vec{k}-\vec{k}^\prime)^2 +k^{\prime 2}(k_x -k_x^\prime)\right\}
\end{eqnarray*}

\begin{eqnarray*}
\Sigma_{\hat{p}_y\hat{P}_y} &=& \int {d^3\vec{k}^\prime\over
(2\pi)^3} {d\omega^\prime\over
2\pi} \left\{G_{\rho\rho}(\vec{k}^\prime,\omega^\prime)G_{\rho\rho}
(\vec{k}-\vec{k}^\prime,\omega-\omega^\prime) \right.\nonumber\\
&&\cdot\left( -{B^2\over 4} k^{\prime 2}(k^\prime _y)^2
(\vec{k}\cdot\vec{k}^\prime)\right.\nonumber\\
&&\left.+I^2 k^\prime_y (k_z -k_z^\prime)^2 (k^\prime_y (k^\prime_z)^2 -
k^\prime_y (k_z -k_z^\prime)^2\right)\nonumber\\
&&-I^2(\lambda+1)\left({1\over 4}(\lambda+1)k_z^2 k^\prime_y (k_z -
k_z^\prime)\right.\nonumber\\
&&-\left.k_z k^\prime_y (k_z -k_z^\prime)^2\right)\cdot
\left(k^\prime_y (k_z -k_z^\prime)-k^\prime_y
k^\prime_z\right)\nonumber\\
&& - \left(I(k^\prime_y k^{\prime 2}_z -k^\prime_y (k_z -
k_z^\prime)^2)\right.\nonumber\\
&& +{I\over 2} (\lambda +1)k_z (k^\prime_y (k_z -k_z^\prime)-
k^\prime_y k^\prime_z))\nonumber\\
&&\left.\cdot Bk^\prime_y (\vec{k}\cdot\vec{k}^\prime )\right)\nonumber\\
&&+G_{n_x n_x}(\vec{k}^\prime,\omega^\prime) G_{n_x n_x} (\vec{k}-
\vec{k}^\prime,\omega-\omega)\nonumber\\
&& \cdot(-K^2 k^\prime_y(\vec{k}-\vec{k}^\prime)^2 (k^\prime_y
(\vec{k}-\vec{k}^\prime)^2 -k^\prime_y k^{\prime 2})\nonumber\\
&&+(G_{n_y n_y} (\vec{k}^\prime,\omega)G_{n_y n_y} (\vec{k} -
\vec{k}^\prime, \omega-\omega^\prime)\nonumber\\
&&\cdot\left(K^2 k^\prime_y (\vec{k}-\vec{k}^\prime)^2 (k^\prime_y (\vec{k}-
\vec{k}^\prime)^2 -k_y^\prime k^{\prime 2})\right)
\end{eqnarray*}

\begin{eqnarray*}
\Sigma_{\hat{p}_z\hat{P}_z} &=& \int {d^3\vec{k}^\prime\over
(2\pi)^3} {d\omega^\prime\over
2\pi} \left\{G_{\rho\rho}(\vec{k}^\prime,\omega^\prime)G_{\rho\rho}
(\vec{k}-\vec{k}^\prime,\omega-\omega^\prime) \right.\nonumber\\
&&\cdot \left({1\over 4}(Ak_z +B(k^\prime_z (\vec{k}\cdot\vec{k}^\prime)-
k_z\vec{k}^\prime\cdot(\vec{k}-\vec{k}))\right)\nonumber\\
&&\cdot (2Ak_z +B (k^\prime_z k^{\prime 2}+k_zk\vec{k}^\prime\cdot
(\vec{k}^\prime-\vec{k}))\nonumber\\
&&-I^2 (k_z -k_z^\prime)^3 ((k_z-k_z^\prime)^3 +k_z^{\prime 3})\nonumber\\
&&-I^2(\lambda-1)(k_z-k_z^\prime)^3 k_x (k^2_z -k_z^{\prime
2})\nonumber\\
&&-I((k_z-k^\prime_z)^3)+k^{\prime 3}_z)\nonumber\\
&&\cdot(Ak_z +B(k^\prime_z(\vec{k}-\vec{k}^\prime)-k_z\vec{k}^\prime
\cdot(\vec{k}-\vec{k}\prime))\nonumber\\
&&+\left(G_{n_x n_x}(\vec{k}^\prime,\omega)G_{n_x n_x}(\vec{k}-
\vec{k}^\prime,\omega-\omega^\prime)+\right.\nonumber\\
&&\left.G_{n_y n_y}(\vec{k}^\prime\omega^\prime)G_{n_y n_y}(\vec{k}-
\vec{k}^\prime, \omega-\omega^\prime)\right)\nonumber\\
&&\cdot(-K^2)\left(k^\prime_z (\vec{k}-\vec{k}^\prime)^2 (k_z^\prime
(\vec{k}-\vec{k})^2\right.\nonumber\\
&&+\left.k^{\prime 2}(k_z -k_z^\prime)\right) +(k^{\prime 2}+(\vec{k}-
\vec{k}^\prime)^2).\nonumber\\
&&\left.(\vec{k}-\vec{k}^\prime)^2 (\lambda^2 k^2_z -2\lambda k_z
k^\prime_z)\right) \nonumber\\
\end{eqnarray*}

\begin{eqnarray*}
\Sigma_{\hat{P}_x\hat{P}_z} &=&  \int {d^3\vec{k}^\prime\over
(2\pi)^3} {d\omega^\prime\over
2\pi} \left\{G_{\rho\rho}(\vec{k}^\prime,\omega^\prime)G_{\rho\rho}
(\vec{k}-\vec{k}^\prime,\omega-\omega^\prime) \right.\nonumber\\
&\cdot&\left(Ak_x +B(k^\prime_x (\vec{k}\cdot\vec{k}^\prime)-
k_x\vec{k}^\prime\cdot(\vec{k}-\vec{k}^\prime))\right)\nonumber\\
&&\cdot\left(2Ak_z +B(k^\prime_zk^{\prime
2}+k_z\vec{k}^\prime\cdot(\vec{k}^\prime -\vec{k}))\right)\nonumber\\
&&-{I\over 2}(Ak_x +B(k^\prime_x (\vec{k}\cdot\vec{k}^\prime)-
k_x\vec{k}^\prime\cdot(\vec{k}-\vec{k}^\prime))\nonumber\\
&&((k_z -k_z^\prime)^3+k_z^{\prime 3})+(\lambda-1)k_x k_z^\prime (k_z -
k_z^\prime))\nonumber\\
&&-I^2 (k_x-k_x^\prime) (k_z-k_z^\prime)^2 ((k_z-k_z^\prime)^3
+k_z^{\prime 3})\nonumber\\
&& -{I^2\over 2}(\lambda +1)k_z (k_z -k_z^\prime)^3 (k_x^\prime
(k_z -k_z^\prime)+k_z^\prime (k_x -k_x^\prime))\nonumber\\
&&I^2(\lambda -1)k_z k_z^\prime (k_x -k_x^\prime)(k_z -
k_z^\prime)^3\nonumber\\
&&-{I^2\over 2}(\lambda^2 -1)k_z k_z k^\prime_x k^\prime_z (k_z -
k^\prime_z)\nonumber\\
&&-{I\over 2}(\lambda +1)k_z k^\prime_z (k_x-k_x^\prime).\nonumber\\
&&(Ak_z+B(k^\prime_z (\vec{k}\cdot\vec{k}^\prime)-
k_z\vec{k}^\prime\cdot(\vec{k}-\vec{k}^\prime)\nonumber\\
&&+G_{n_x n_x}(\vec{k}^\prime,\omega^\prime)G_{n_x n_x} (\vec{k}-
\vec{k}^\prime,\omega-\omega^\prime)\nonumber\\
&&\cdot(\vec{k}-\vec{k}^\prime)^2 K^2 (-{1\over 2}(\lambda +1)k_x
(k_z^\prime(\vec{k}-\vec{k}^\prime)^2 +k^{\prime 2}(k_z -
k_z^\prime))\nonumber\\
&&-{1\over 2}(\lambda -1)k_x k_z^\prime (k^{\prime 2}+(\vec{k}-
\vec{k}^\prime)^2))\nonumber\\
&&+\lambda^2 K^2k_xk_z(\vec{k}-\vec{k}^\prime)^2(k^{\prime
2}+(\vec{k}-\vec{k}^\prime)^2)\nonumber\\
&&+(G_{n_x n_x}(\vec{k}\prime,\omega^\prime)G_{n_x n_x}(\vec{k}-
\vec{k}^\prime,\omega-\omega^\prime))\nonumber\\
&&+G_{n_y n_y}(\vec{k}^\prime,\omega^\prime)G_{n_y n_y}(\vec{k}-
\vec{k}^\prime,\omega-\omega^\prime))\nonumber\\
&&\cdot(-K^2) (\vec{k}-\vec{k}^\prime)^2 k_x^\prime\nonumber\\
&&\left(k_z^\prime(\vec{k}-\vec{k}^\prime)^2+k^{\prime 2}(k_z -
k_z^\prime\right)\nonumber\\
&&\left. +k_z k^{\prime 2}+(\vec{k} -
\vec{k}^\prime)^2k_z\right)\nonumber\\
\end{eqnarray*}

\begin{eqnarray*}
\Sigma_{\hat{n}_x \hat{n}_x}&=&  \int {d^3\vec{k}^\prime\over
(2\pi)^3} {d\omega^\prime\over
2\pi} \left\{G_{\rho\rho}(\vec{k}^\prime,\omega^\prime)G_{\rho\rho}
(\vec{k}-\vec{k}^\prime,\omega-\omega^\prime) \right.\nonumber\\
&&\cdot ({I\over \gamma_1})^2 k^\prime_x (k_z-k_z^\prime)(k^\prime_x
(k^\prime_z-k_z)+k^\prime_z (k^\prime_x-k_x))
\end{eqnarray*}

The remaining self-energies needed, $\Sigma_{n_x \hat{n}x},
\Sigma_{\hat{n}_y P_y}$ and $\Sigma_{\hat{P}_x\rho}$, do not exhibit
feedback. The first of these renormalize the Frank constants and has
no graphs of the form we are considering. However, the finite graph
contributing to $\Sigma_{n_x \hat{n}_x}$ does break the one-constant
approximation. Similarly, the other two self-energies also do not have
any graphical contributions of the form we are considering.
\par
With the self-energies (A1) -- (A5) the viscosities $\nu_i, i=1,\ldots
5$ can principle be calculated to one-loop order using (3.32) --
(3.36). In general these expressions are quite complicated and not
much will be learned by displaying them. Even the hydrodynamic limit
is difficult to evaluate (except for $\nu_2$ and $\nu_3$) due to the
anistropy of the propagators. Finally the functions $f_0, f_1, f_2,
f_3, g_0, g_1, g_2, h_1$ and $h_2$ appearing in (3.53) -- (3.55) can
in principle be calculated using (3.37) -- (3.39) and (A1) -- (A5).
\vfil\eject

\centerline{\bf References}
\noindent
$^*$ Present address: \ Spectral Sciences, Inc., 99 S. Bedford St.,
Burlington, MA 08103

\begin{enumerate}
\item For experimental studies of a nematic glass, see e.g. J. I.
Spielberg and E. Gelerinter, Phys. Rev. A. {\bf 32} , 3647  (1985), H.
R. Zeller, Phys. Rev. Lett. {\bf 48}, 334 (1982), and W. J. Laprice
and D. L. Uhrich, J. Chem. Phys. {\bf 71}, 1498 (1979).
\item M. Gabay and G. Toulouse, Phys. Rev. Lett. {\bf 47}, 201 (1981).
\item For a theoretical study for an anisotropic glassy system see e.g.
L. Golubovic and T. C. Lubensky, Phys. Rev. Lett. {\bf 63}, 1082
(1989).
\item E. Leutheusser, Phys. Rev. A{\bf 29}, 2765 (1984). For a review
see W. G\"{o}tze, in  ``Liquids, Freezing and Glass Transitions", J.
P. Hansen, D. Levesque and J. Zinn-Justin, eds. North-Holland, New
York, 1991.
\item S. Das, G. Mazenko, S. Ramaswamy and J. Toner, Phys. Rev.  Lett.
{\bf 54}, 118 (1985). For a review see B. Kim and G. F. Mazenko,
Advances in Chemical Physics, {\bf 78}, 129 (1990).
\item S. Das and G. Mazenko, Phys. Rev. A {\bf 34}, 2265 (1986).
\item R. Schmitz, J. W. Dufty and P. De, unpublished.
\item See the review articles cited in refs. 4 and 5.
\item B. Kim, Phys. Rev. A {\bf 46}, 1991 (1991).
\item See e.g. Y. Y. Goldschmidt in ``Recent Progress in Random
Magnets", D. H. Ryan, ed., World Scientific Singapore, 1992.
\item For an introduction to liquid crystals, see, e.g., P.G. de
Gennes,   ``The Physics of Liquid Crystals", Clarendon, Oxford, 1974.
\item See refs. 11, 12, and 16, as well as M. J. Stephen and J. P.
Straley, Rev. Mod. Phys. {\bf 46}, 617 (1974), D. Forster, T. C.
Lubensky, P. C. Martin, J. Swift and P. S. Pershan, Phys. Rev. Lett.
{\bf 26}, 1016 (1971), F. J\"{a}hnig and H. Schmidt, Ann. Phys. {\bf
71}, 129 (1972).
 \item P. C. Martin, O. Parodi and P. S. Pershan, Phys. Rev. A. {\bf
6}, 2401 (1972).
\item S. Ma and G. F. Mazenko, Phys. Rev. B {\bf 11}, 4077 (1975).
\item We have found that the nonlinearities in $\delta n_i$ do not
yield a feedback mechanism and in the interest of simplicity we will
ignore them from the start.
\item D. Forster, Annals. of Phys. {\bf 85}, 505 (1974) and
``Hydrodynamic Fluctuations, Broken Symmetry and    Correlation
Functions", W. A. Benjamin, (1975).
\item D. Foster, Phys. Rev. Lett. {\bf 32}, 1161 (1974).
\item S. Das, Phys. Rev. A {\bf 36}, 211 (1987).
\item P. C.   Martin, E. D. Siggia and H. A. Rose, Phys. Rev. A. {\bf
8}, 423 (1973).
\item S. D. Goren, C. Korn and S. B. Marks, Phys. Rev. Lett. {\bf 34},
1212 (1975); J. W. Doane, C. E. Tarr and M. A. Nickerson, Phys. Rev.
Lett. {\bf 33}, 620 (1974); S. Meiboom and R. C. Hewitt, Phys. Rev.
Lett. {\bf 30}, 261 (1973).
\item S. Das, Phys. Rev. A. {\bf 42}, 6116 (1990).
\end{enumerate}

\end{document}